\documentclass{revtex4} 
\usepackage{epsfig}

\begin{document}
\title{Comments on recent work on dark-matter capture in the Solar System}

\author{Joakim Edsj\"o}
\email{edsjo@fysik.su.se}
\affiliation{The Oskar Klein Centre for Cosmoparticle Physics, Department of Physics, Stockholm University, AlbaNova, SE - 106 91 Stockholm, Sweden}
\author{Annika H. G. Peter}
\email{apeter@astro.caltech.edu}
\affiliation{California Institute of Technology, MC 249-17, Pasadena, CA 91125}

\begin{abstract}
Recently, several papers have appeared that examine the process of capturing dark-matter particles from the Galactic halo to orbits bound to the Solar System. The authors of these papers predict large enhancements to the local dark-matter density via gravitational three-body interactions with planets. However, these conclusions are wrong; these papers do not include the inverse process to capture, namely the ejection of dark-matter particles by three-body gravitational encounters.  We emphasize previous work that shows that by including both capture and ejection of dark matter from the Solar System, the density of dark matter bound to the Solar System is small compared to the local Galactic dark-matter density.
\end{abstract}

\maketitle



Dark-matter particles in the Galactic halo can end up on orbits gravitationally bound to the Solar System via gravitational three-body interactions with the planets \cite{Gould-diff}.  As a consequence, the Solar System contains a population of dark-matter particles bound to it, in addition to the Galactic population that streams through the system.  While most previous literature on the subject has found the bound population to be small \cite{Gould-diff,Lundberg:2004dn,Peter2009a,Peter:2009mm}, three papers have appeared recently by Xu \& Siegel \cite{Xu:2008ep}, Khriplovich \& Shepelyansky \cite{Khriplovich:2009jz}, and Khriplovich \cite{Khriplovich:2010hn} in which the authors find huge enhancements of the density of dark matter bound to the Solar System compared to the local dark-matter density in the Galactic halo. There has also appeared several papers (see e.g.\ \cite{Cerdonio:2008un,Iorio:2010gh,Adler:2008vc}) that use these huge enhancements.

Even if the techniques used in these recent papers differ from previous studies and from each other, of which some could be developed to become useful techniques for other Solar System studies, they all share a common flaw in that they incorrectly neglect the inverse process of capture from the Galactic halo: ejection of dark-matter particles from the Solar System.  

In the limit of a short-duration interaction with a planet (relative to the orbital time scale of the planet), the three-body interactions of the Sun + planet + dark-matter particle system go as follows.  A dark-matter particle approaching a planet is deflected in the gravitational field of the planet, although the speed of the particle with respect to the planet is unchanged in the center-of-mass frame of the particle-planet system.  While the velocity with respect to the planet is unchanged in this interaction, the velocity with respect to the Solar System has changed.  Hence, if a Galactic dark matter particle ends up with a velocity less than the escape velocity from the Sun it is bound (captured) to the Solar System.

Of course, this deflection also occurs for dark-matter particles already captured in the Solar System, and the dark-matter particles run a risk of being ejected from the Solar System.  If we imagine that the Solar System started out devoid of bound dark matter at its birth, the capture process would initially dominate.  After a while, though, when the density of captured dark-matter particles gets high enough, the inverse process of ejection will start competing.  Eventually, the two processes will reach equilibrium, and hence the density of dark matter in the Solar System will be fixed.   It can be shown by detailed balance that the equilibrium density of bound dark-matter particles f($v$) (at a speed $v$ in the Solar System) is approximately the same as the corresponding phase-space density $f(u)$ (if $u$ is a heliocentric speed of a dark-matter particle well outside the sphere of influence of the Sun) in the Galactic halo, well outside the gravitational potential well of the Solar System, if $u$ is replaced by $v$ in the phase-space density.  This was first shown by Gould in 1991 \cite{Gould-diff}. Since only the low-velocity part of the Galactic halo dark-matter population is subject to capture, the local spatial density of captured dark-matter particles is at most only a fraction of the Galactic halo density of dark matter.

With detailed calculations, one can calculate the timescale for equilibrium to occur.  For some parts of phase space, equilibrium will occur on time scales shorter than the age of the Solar System.  For other parts, the time scale is longer and the phase-space density will be even lower than the equilibrium density (see e.g.\ Fig.~3 in \cite{Gould-diff} and Sec.~III in \cite{Peter:2009mm}).

The analytical calculations by Gould in 1991 \cite{Gould-diff} have been supplemented with numerical simulations by Lundberg \& Edsj\"o \cite{Lundberg:2004dn} and by Peter \cite{Peter:2009mm}, the latter of which also explores the regime of long-duration encounters of halo dark-matter particles with the Solar System.  In both these numerical simulations, the analytical arguments by Gould are confirmed, detailed balance does hold and we get a limiting density of dark matter particles in the Solar System that is much lower than found in the papers by Xu \& Siegel \cite{Xu:2008ep}, Khriplovich \& Shepelyansky \cite{Khriplovich:2009jz}, and Khriplovich \cite{Khriplovich:2010hn}.

The short lifetimes of small objects in the Solar System is well established in the field of planetary science.  A range of numerical studies show that planetesimals, comets or asteroids on outer-planet-crossing orbits are ejected from the outer Solar System on $\sim \hbox{Myr}$ timescales \cite{duncan1987,gladman1990,dones1999,malyshkin1999}.  Planetesimals or asteroids originating in the inner Solar System are ejected or driven into the Sun or planets on $\sim 10\hbox{ Myr}$ time scales \cite{farinella,gladman2000}.  If ejection were truly unimportant, the Solar System would be littered with debris from the original proto-planetary disk.

The only case in which the population of dark-matter particles bound to the Solar System could be increased is if processes other than pure gravitational diffusion could be at play, e.g.\ weak interactions with atoms in the Sun or the planets, which was studied by Damour \& Krauss \cite{Damour:1998rh}. For this mechanism to significantly boost the bound dark-matter population requires that a significant fraction of the particles captured by weak-scale interactions with atoms in the Sun or planets have long lifetimes in the Solar System.  
This can happen if there are resonances that prolong a number of particle lifetimes in the Solar System by pulling the orbital perihelia outside the Sun for extended times.  Otherwise, dark-matter particles quickly lose all their energy to solar nuclei with a typical time scale of the initial orbital period divided by the optical depth of weak-scale interactions in the Sun, form a dense core in the Sun, and annihilate.  Dark-matter particles captured in the Solar System by weak-scale interactions with solar nuclei were investigated by Peter \cite{Peter2009a} in the context of a simplified solar system consisting only of Jupiter and the Sun.  It was found that although there was a small population of bound particles whose lifetimes were extended well past the typical time scale, the overall effect on the bound dark-matter density was small.  However, this was not the scenario envisioned by Xu \& Siegel \cite{Xu:2008ep}, Khriplovich \& Shepelyansky \cite{Khriplovich:2009jz}, and Khriplovich \cite{Khriplovich:2010hn}.

In conclusion, we emphasize that any calculation of the effect of a process on an observable should include the inverse process.  Neglecting the ejection of dark matter from the Solar System, as is done in \cite{Xu:2008ep,Khriplovich:2009jz,Khriplovich:2010hn}, leads to estimates of the bound population of dark matter in the Solar System that are many orders of magnitude too large. In contrast, calculations that do include the ejection \cite{Gould-diff,Lundberg:2004dn,Peter2009a,Peter:2009mm} find much more modest densities of dark matter in the solar system.

\section*{Acknowledgements}

J.E.~thanks the Swedish Research Council (VR) for support. A.P.~thanks the Gordon and Betty Moore foundation for support.

\end{document}